\documentclass{iaus}
\usepackage{graphicx}
\title{Probe of dark galaxies via
 disturbed/ lopsided isolated galaxies}

\author{Igor D. Karachentsev$^1$,
 Valentina E. Karachentseva$^2$,
 Walter K. Huchtmeier$^3$,
 Dmitry I. Makarov$^1$,
and Serafim S. Kaisin$^1$}

\affiliation{$^1$Special Astrophysical Observatory, Russian Academy of Sciences,
	 N.Arkhyz, KChR, 369167, Russia \break email:ikar@sao.ru,dim@sao.ru,
 skai@sao.ru\\
$^2$Astronomical Observatory of Kiev University, Observatorna 3, 04053
       Kiev, Ukraine, \break email:vkarach@observ.univ.kiev.ua\\
$^3$Max-Planck-Institut fur Radioastronomy, Auf dem Hugel 69,
    D-53121 Bonn, Germany \break email:p083huc@mpifr-bonn.mpg.de}

\pubyear{2007}
\volume{244}  
\pagerange{}
\date{?? and in revised form ??}
\jname{Proceedings Title IAU Symposium}
\editors{A.C. Editor, B.D. Editor \& C.E. Editor, eds.}
\begin{document}

\maketitle

\begin{abstract}
    Searching for lopsided/interacting objects among ~1500 isolated
galaxies yields only eight strongly disturbed galaxies which may be
explained as a result of their interaction with massive dark objects.
We present results of spectral and photometric observations of these
galaxies performed with the 6-m telescope that lead to significant
restriction on cosmic abundance of dark galaxies.
\keywords{dark matter, interaction galaxies, peculiar galaxies}
\end{abstract}

\section{Introduction}
   The standard Lambda-CDM cosmology assumes that besides dark halos
with luminous galaxies in their cores, completely dark clumps
(sub-halos) should also exist with masses $10^8 - 10^{11}M_{\odot}$
(\cite[van den Bosch et al. 2003]{Bosh03}, \cite[Tully 2005]{Tul05},
\cite[Yang et al. 2005]{Yang05}). The total
number of dark sub-halos may exceed in the number of the usual galaxies
by a factor tens. Within the Local Group, \cite[Klypin et al. (1999)]{Kly99} predict
about 300 satellites with masses greater than $~3*10^8M_{\odot}$
that is significantly higher than the observed one.

   A possible explanation for this discrepancy is the physical
processes inhibiting star formation especially in low mass clumps,
thus implying the existence of a large number of dark satellites.

Here, we use an approach proposed by \cite[van den Bergh (1969)]{Berg69} and \cite[Trentham
et al. (2001)]{Tren01}, which relies on searching for signs of non-motivated
distortion visible on images of spatially isolated galaxies.

  As known, galaxies in close encounters show a significant signature
of interaction in the form of a distortion of their structure,
the presence of tails and bridges, or a common diffuse envelope.
All these features have been quantitatively explained based on numerous
N-body simulations ever since \cite[Toomre \& Toomre (1972)]{Tom72}.

  Signs of interaction are seen in more than 50\% of those binary galaxies
where the separation is comparable to the sum of the diameters. In systems
with greater separation, as in triplets, the relative number of peculiar
galaxies is about 30\%. In magnitude-limited galaxy catalogs, the fraction
of interacting objects is equal to 8\% (\cite[Karachentsev,``Binary galaxies'', 1987]{Karach87}).

  When considering more and more scattered systems of galaxies and
single galaxies of the ``general field'', one can expect a fraction
of the interacting objects among them to be nearby zero. This would
occur if there were no other objects except the luminous galaxies.

  However, if completely dark galaxies with masses of $10^8 - 10^{11}M_{\odot}$
exist, the phenomena of interaction will occur in the case of
extremely isolated galaxies too.  Hence, an asymptotic relative
number of peculiar shapes among the most spatially isolated galaxies
may be a sensitive tool for estimating the cosmic abundance of massive
dark galaxies.

\section{Searching for peculiar isolated galaxies}

  To search for such strange cases of interaction where the second
interacting companion is invisible, we used the ``Catalog of Isolated
Galaxies'' (\cite[Karachentseva 1973]{Karach73} = KIG). This catalog contains 1050 galaxies
without ``significant'' neighbors.

  According to our estimates, the catalog objects do not suffer
essential perturbations from neighboring galaxies over some Gyrs.
As it is a sample of northern galaxies with $m < 15.7$, the catalog
includes only 4\% of the CGCG galaxies from \cite[Zwicky et al. (1961-1968)]{Zw61};
hence KIG is a collection of a rather rare kind of galaxies.

  All KIG galaxies were inspected by us on the Digital Sky Survey
In some cases, we also studied galaxy images in the 2MASS to check
for the possibility of double-nuclei systems as recent merging
remnants. The results of our inspections (\cite[Karachentsev et al. 2006])
{Karach06} yield five disturbed/lopsided galaxies.

 Then, we undertook a new search for isolated distorted galaxies
in a sample limited by radial velocities $V_{LG} < 3200$ km s$^{-1}$. Among
7500 such galaxies of the Local Supercluster, about 60\% reside
in groups of different populations (\cite[Makarov \& Karachentsev, 2000]{Mak00}).
The remaining $N ~ 3000$ galaxies are characterized by different
degrees of isolation with respect to their neighbors.

  We selected  500 of the most isolated galaxies and inspected their
images on DSS. Only 4 galaxies out of 500 show significant signs of
interaction. One case, UGC~4722, turns out to be common with
the KIG sample.

  Signs of interaction between galaxies are  best developed when
the objects have similar masses. If the number of completely dark
sub-halos with typical masses of $10^8 - 10^{11}M_{\odot}$ in any volume
approximately corresponds to the number of luminous galaxies,
then one may expect about 8\% interacting galaxies among the isolated
ones (interaction with an invisible object).  This rough estimate
ignores, of course, properties of spatial distribution of dark
sub-haloes with respect to the luminous galaxies.

  The observed relative number of KIG galaxies with clear features
of interaction, 5/1050 = 0.5\%, turns out to be one order of
magnitude lower than expected.

  The observed frequency of disturbed shapes among very isolated
galaxies in the Local Supercluster, 4/500 = 0.8\%, is consistent
with the previous estimate made for the KIG sample.
In view of these results, it is rather unlikely that the number of
massive dark sub-halos is similar to the number of the usual galaxies.

    Moreover, apart from interaction with a dark galaxy, the observed
morphological irregularities of isolated galaxies may different origins,
in particular, there could have been a merger, with the companion
now merged and not visible anymore, or there could be large gas
accretion from cosmic filaments, leading to perturbed morphologies
via star bursts.

 As noted by Trentham, dark galaxies probably have low masses and
negligible dynamical effects on massive galaxies like the ones in the
KIG sample. They can have much more substantial effects on the nearby
low-mass galaxies seen, say, in the Catalog of Neighboring Galaxies
(\cite[Karachentsev et al. 2004]{Karach04}).

  This sample contains 197 quite isolated galaxies with a ``tidal index''
$TI < 0$. (A negative $TI$ means that the Keplerian cyclic period of
the galaxy with respect to its main neighboring disturber exceeds
the cosmic Hubble time.) About 90\% of them are low mass dwarfs.

\cite[Pustilnik et al. (2005)]{Pus05} found that an isolated nearby galaxy, DDO~68 with
$M_B = -14.3$ and $TI = -1.6$, looks as a disturbed object with a long curved
tail. If DDO~68 is a single such object in the Local Volume,
it yields a fraction of disturbed isolated galaxies, 1/197 = 0.5\%,
the same as for the KIG catalog.

\section{ Observations with the BTA 6-m telescope}

   Spectral observations were performed by us for some faint galaxies
around  each of the eight peculiar objects. No physical companions to the
considered peculiar galaxies were found, confirming their high isolation.

  The peculiar galaxies were imaged in $B,V,R$ bands, and in the emission
$H_{\alpha}$ line. Long-slit spectra were obtained also across the target
galaxies to study their kinematics. For the observations we used the
multi-mode focal reducer SCORPIO installed at the BTA 6-m telescope.
Figure 1 presents images of 8 the galaxies in different optical bands.
The right hand images are given in larger scale for better seeing of the
central regions.

 Preliminary analysis of the data allows us to separate eight the galaxies
into three categories, as shown in Table 1:

A) Galaxies of high luminosity
with ripples, plumes and loops, caused apparently by a recent merging;
two of them, UGC~5101 and UGC~11905, have been classified as mergers by
\cite[Rothberg \& Joseph (2004)]{Rot04} based on K-band surface photometry.

B) Galaxies
of low luminosity with asymmetric star formation (ESO~539-7 and NGC~244);
they have a quite regular undisturbed periphery and eccentric HII regions.

C) Galaxies (UGC~4722 and ESO~545-5), whose disturbed peculiar shape may
be really originated by a tight encounter with a massive dark object.
Note that both the galaxies have hydrogen mass-to-luminosity ratio,
$M(HI)/L_B$, higher than 1 in solar units. Results of our spectral and
photometric observations for all eight the galaxies will be discussed in
details in a separate paper.
\begin{table}
\caption{Isolated galaxies with disturbed structure}
\begin{tabular}{llcrc} \hline
 Galaxy &    $V_{LG}$   & $M_B$&   log($L_B$) &  log(MHI)\\
\hline
\multicolumn{5}{c}{Probable mergers:}\\

 UGC~5101  &  +11904&   $-$21.1&   10.60 &       --   \\
 UGC~11905 &  + 7737&   $-$20.6&   10.40 &       --    \\
 UGC~11871 &  + 8230&   $-$21.0&   10.56 &      9.68  \\
 F635-2    &+12000: &   $-$20.7  & 10.44   &     --       \\

\multicolumn{5}{c}{Asymmetric SF:}\\

 ESO~539-7  & + 3279 &  $-$18.6  &  9.60    &   9.74    \\
 NGC~244    & + 1021 &  $-$17.0  &  8.96    &   8.73    \\

\multicolumn{5}{c}{Probably disturbed by a dark galaxy:}\\

 UGC~4722   & + 1705  &  $-$16.9 &   8.92   &    9.41  \\
 ESO~545-5  &  + 2329 &  $-$18.9 &   9.72   &    9.89  \\
\hline
\end{tabular}
\end{table}

Finally, we suggest that the existence of dark galaxies can be a probable
reason of signs of interactions observed in some isolated galaxies.
However, a cosmic abundance of dark galaxies is likely less than 1/20
the population of usual luminous galaxies.

{\bf Acknowledgements} This work was supported by DFG--RFBR grant 06--02--04017.

 {}

\newpage
\begin{figure}
\includegraphics{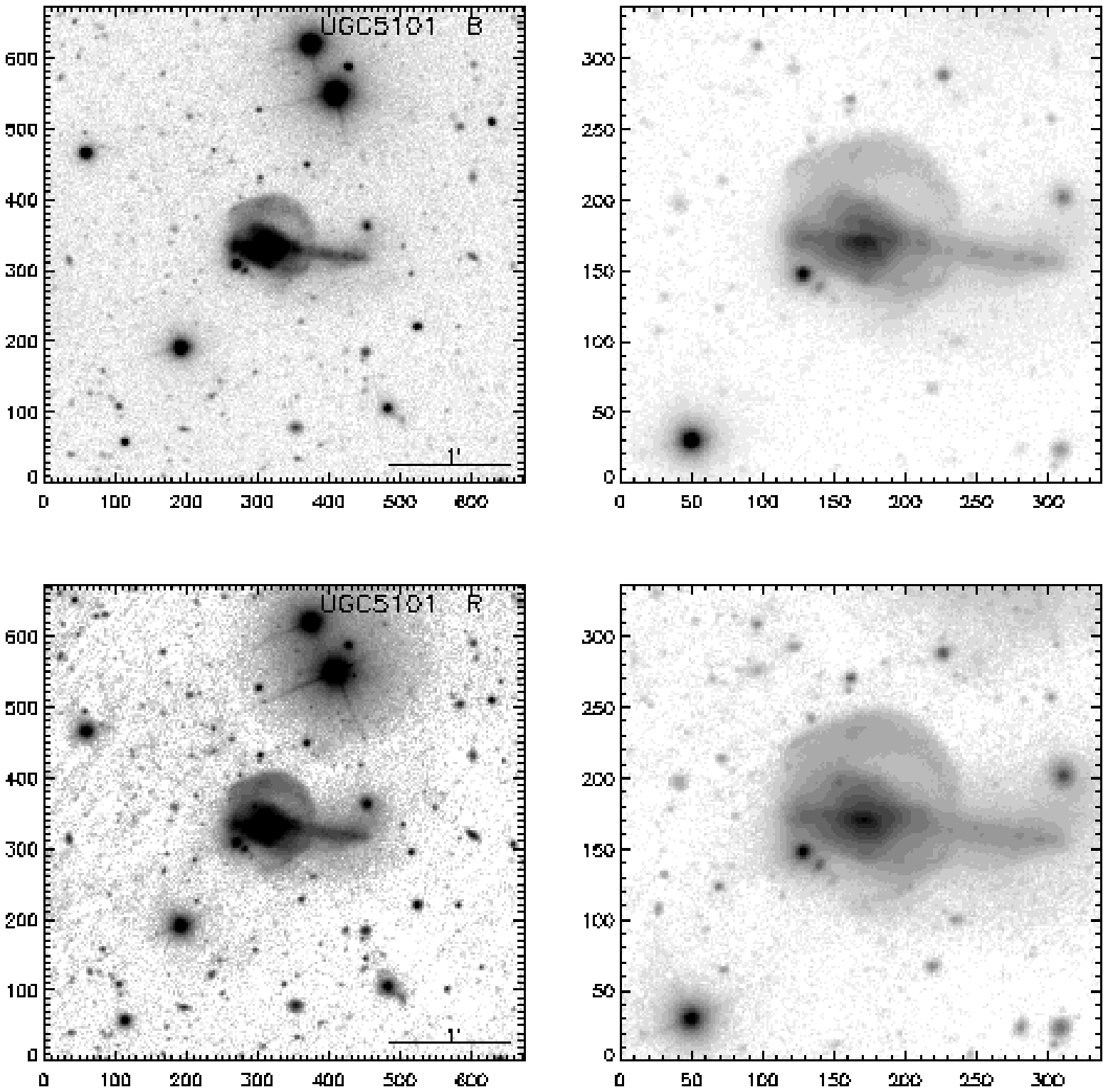}
\caption{BTA images of the eight disturbed isolated galaxies seen in
different optical bands. Enlarged central parts of them are given at
right side. East is left and North is up.}
\end{figure}
\setcounter{figure}{0}
\begin{figure}
\includegraphics{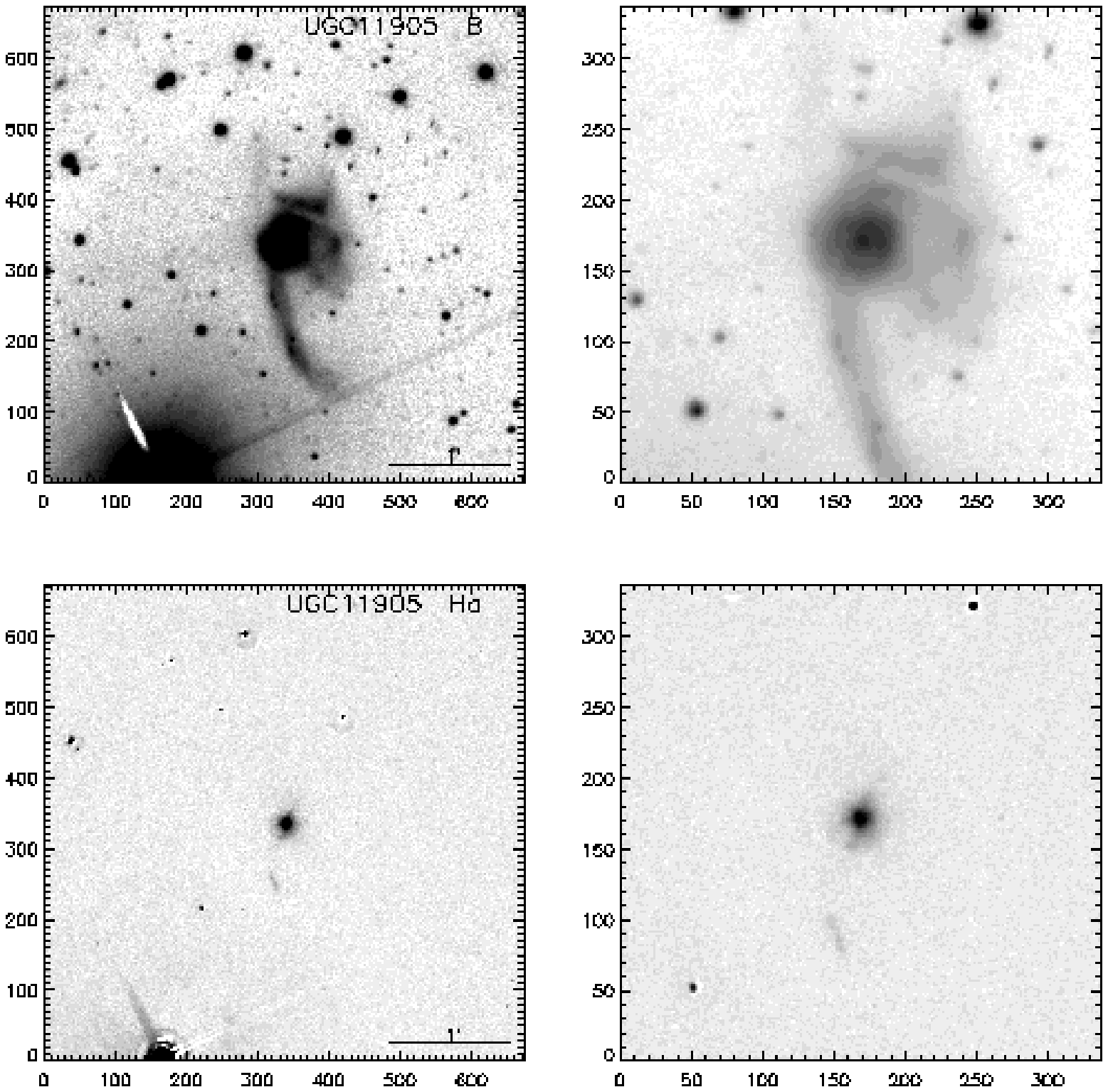}
\caption{Continued}
\end{figure}

\setcounter{figure}{0}
\begin{figure}
\includegraphics{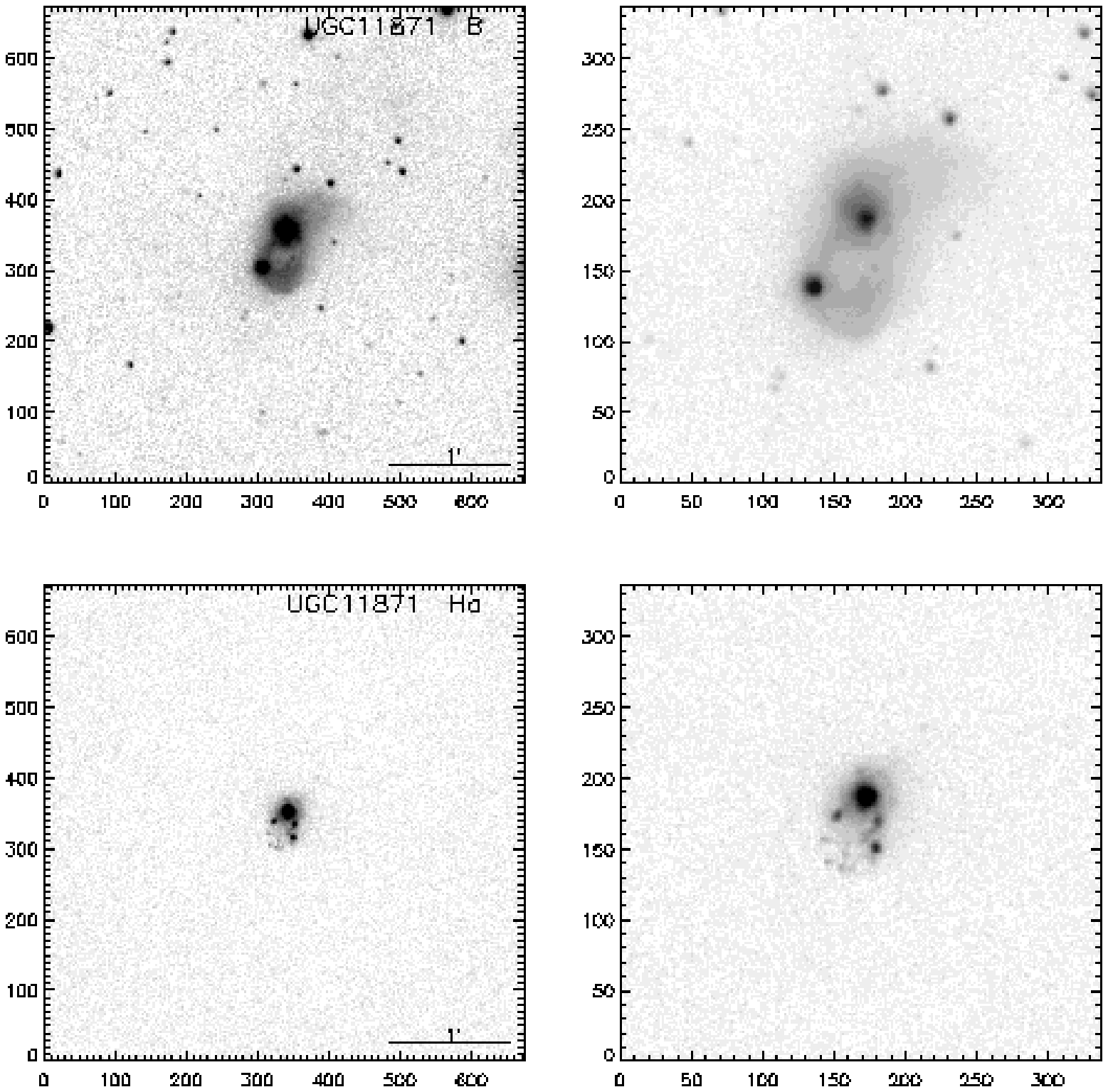}
\caption{Continued}
\end{figure}

\setcounter{figure}{0}
\begin{figure}
\includegraphics{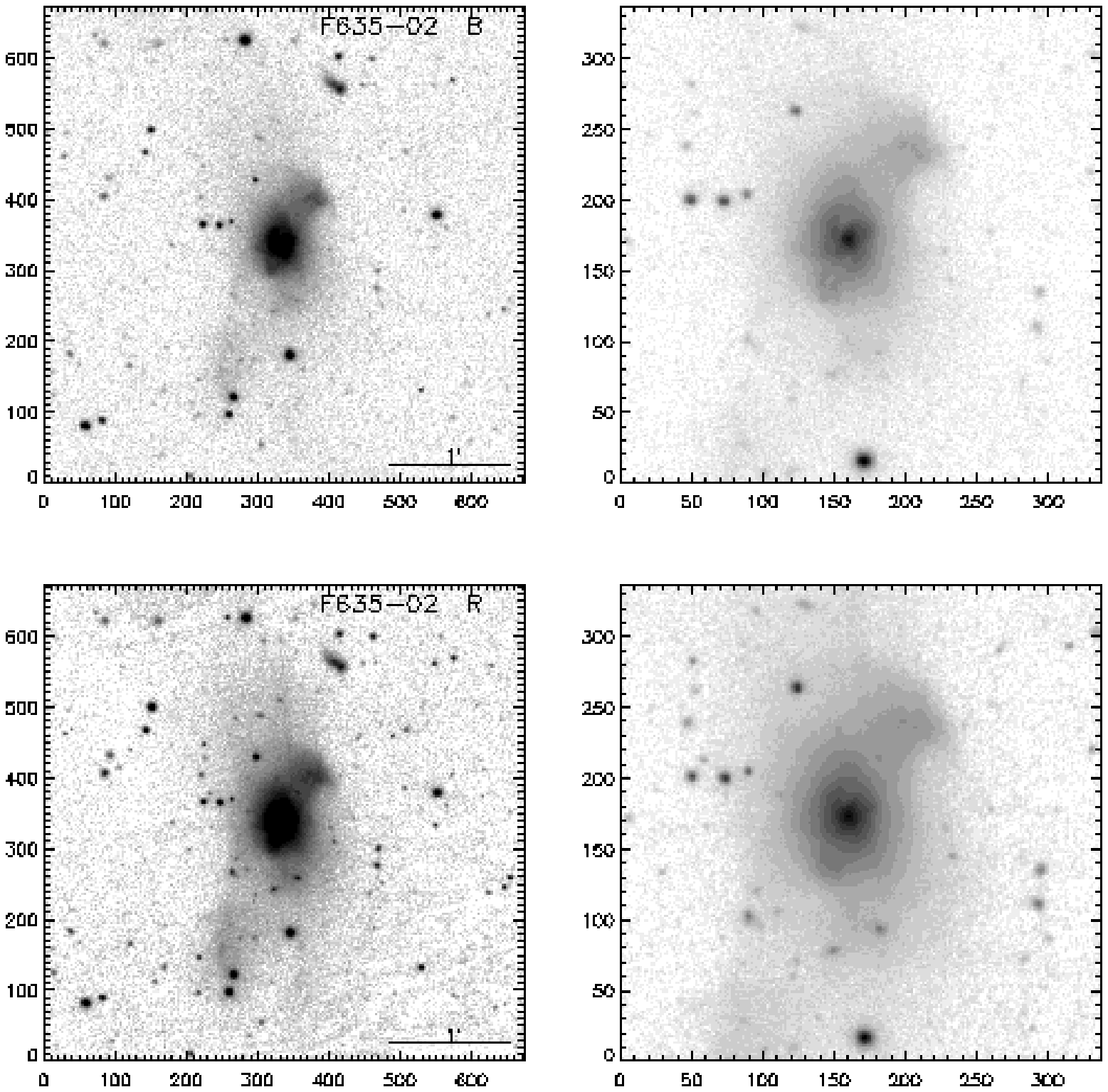}
\caption{Continued}
\end{figure}

\setcounter{figure}{0}
\begin{figure}
\includegraphics{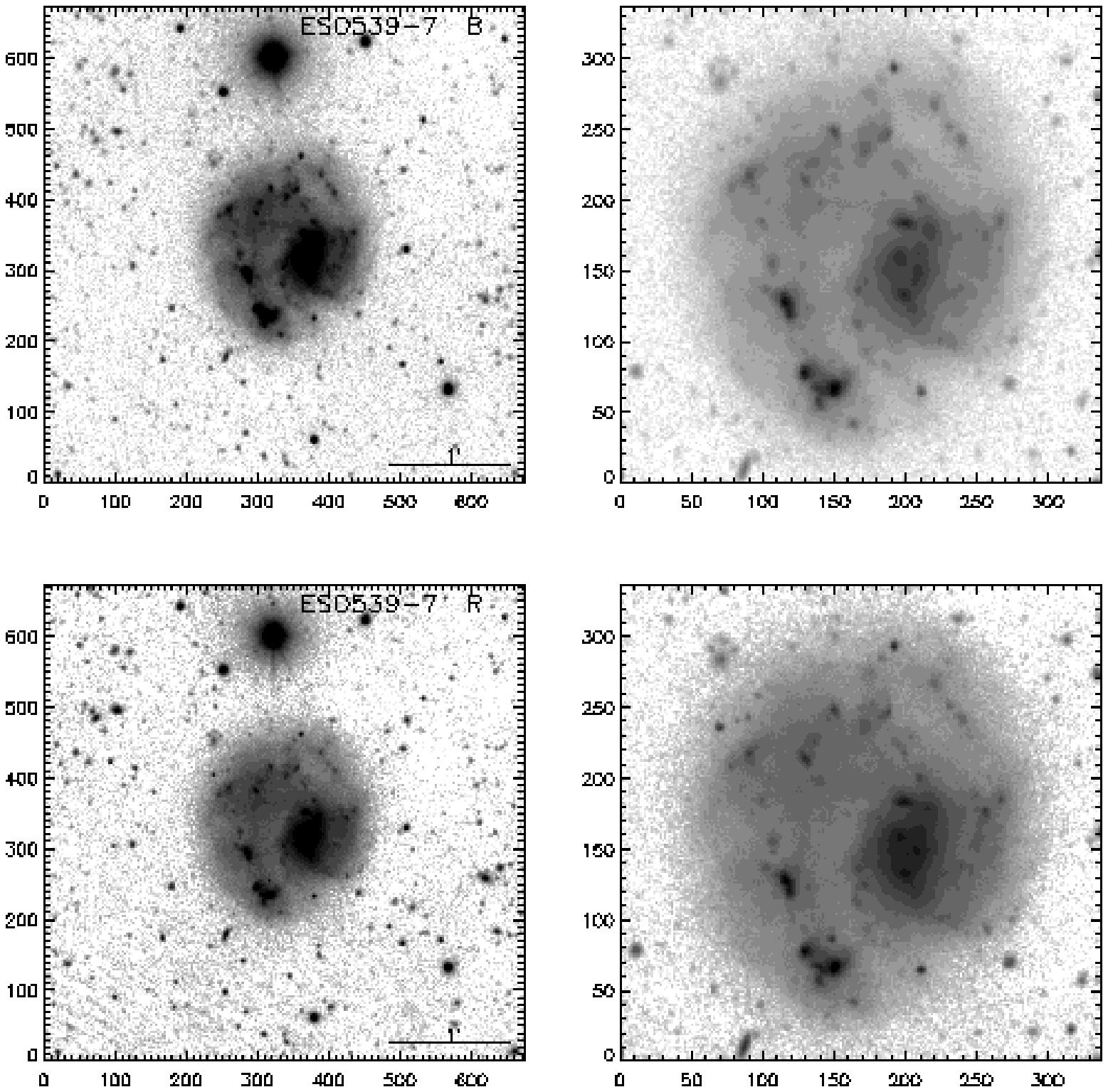}
\caption{Continued}
\end{figure}

\setcounter{figure}{0}
\begin{figure}
\includegraphics{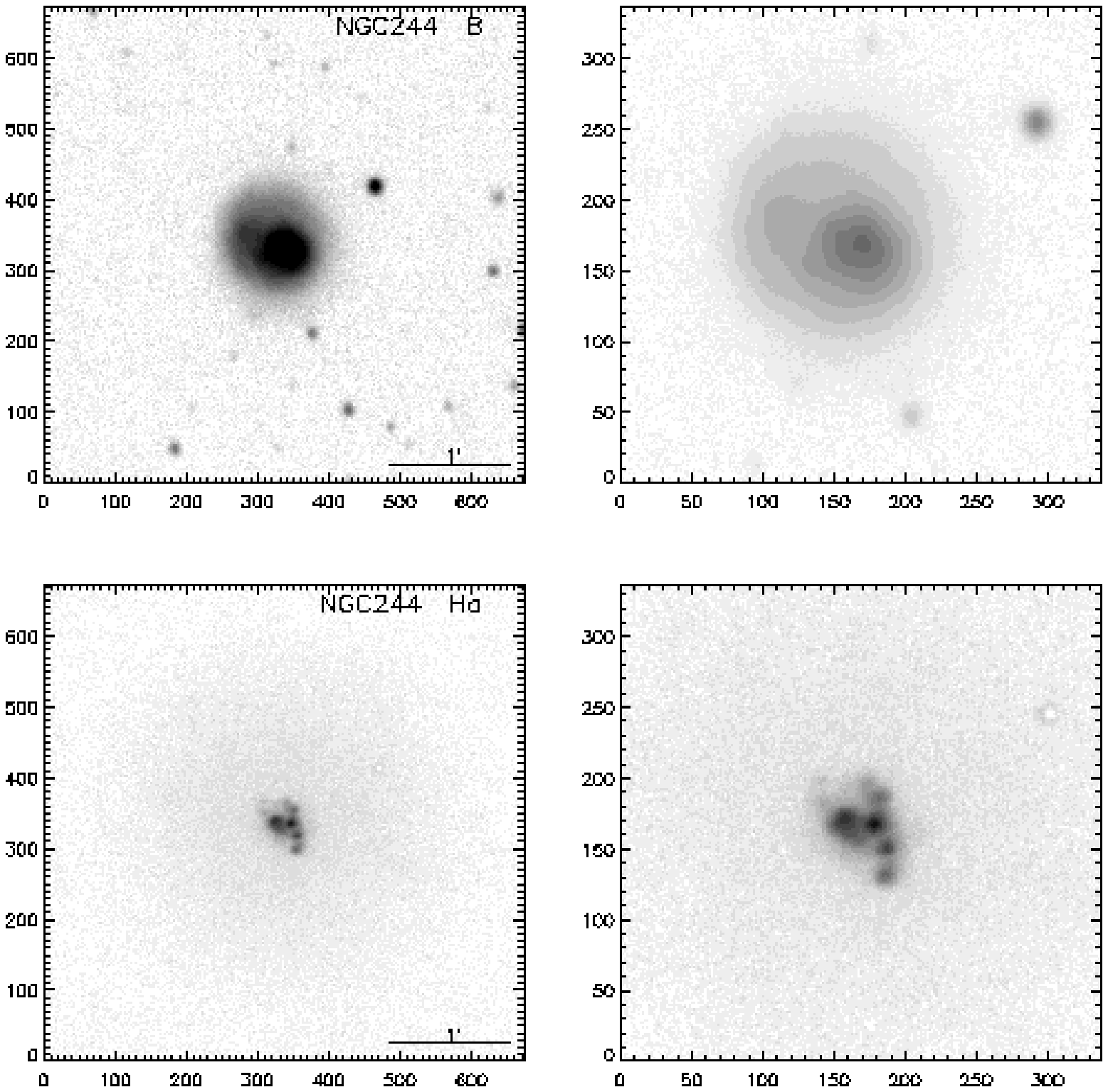}
\caption{Continued}
\end{figure}

\setcounter{figure}{0}
\begin{figure}
\includegraphics{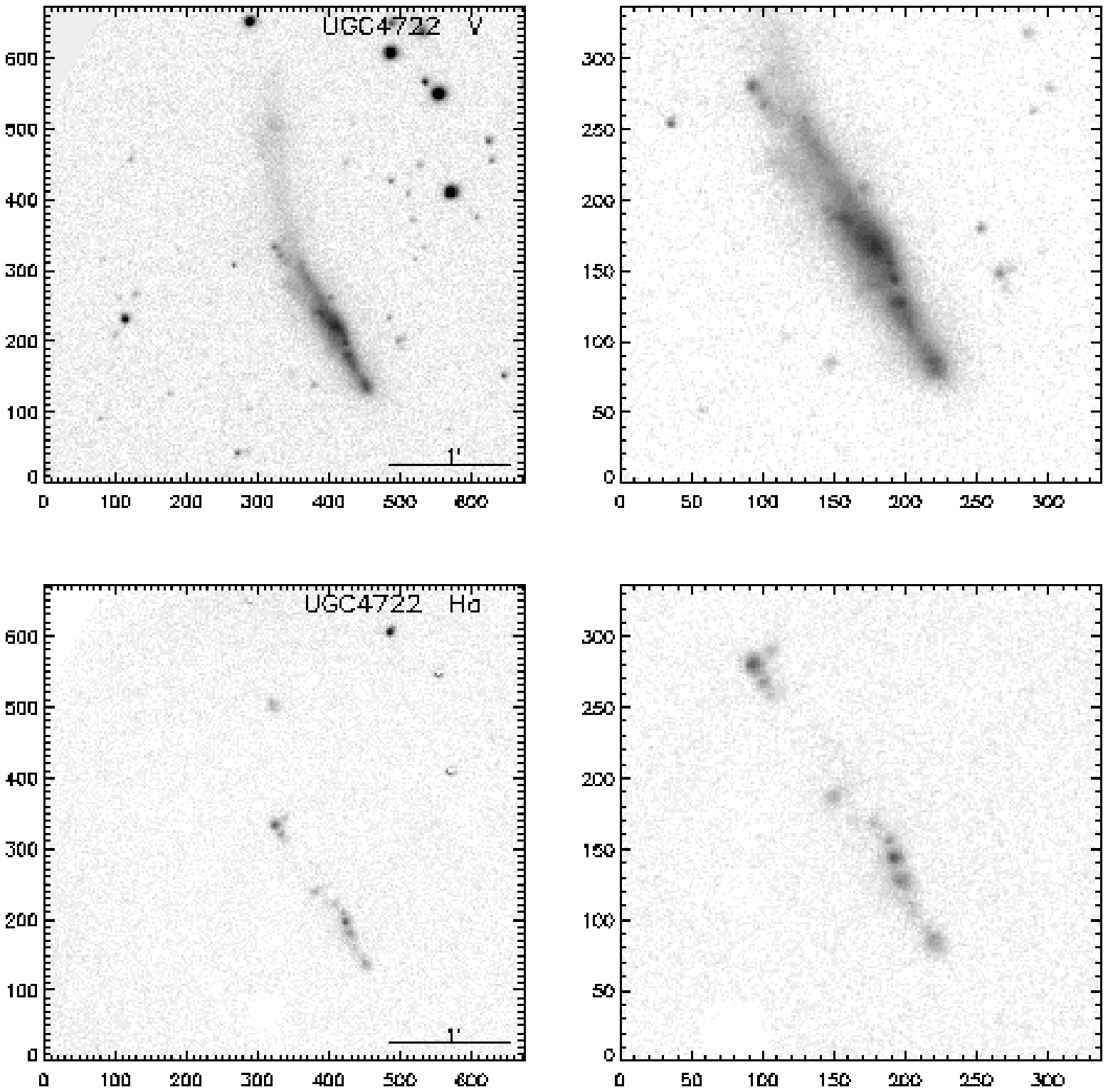}
\caption{Continued}
\end{figure}

\setcounter{figure}{0}
\begin{figure}
\includegraphics{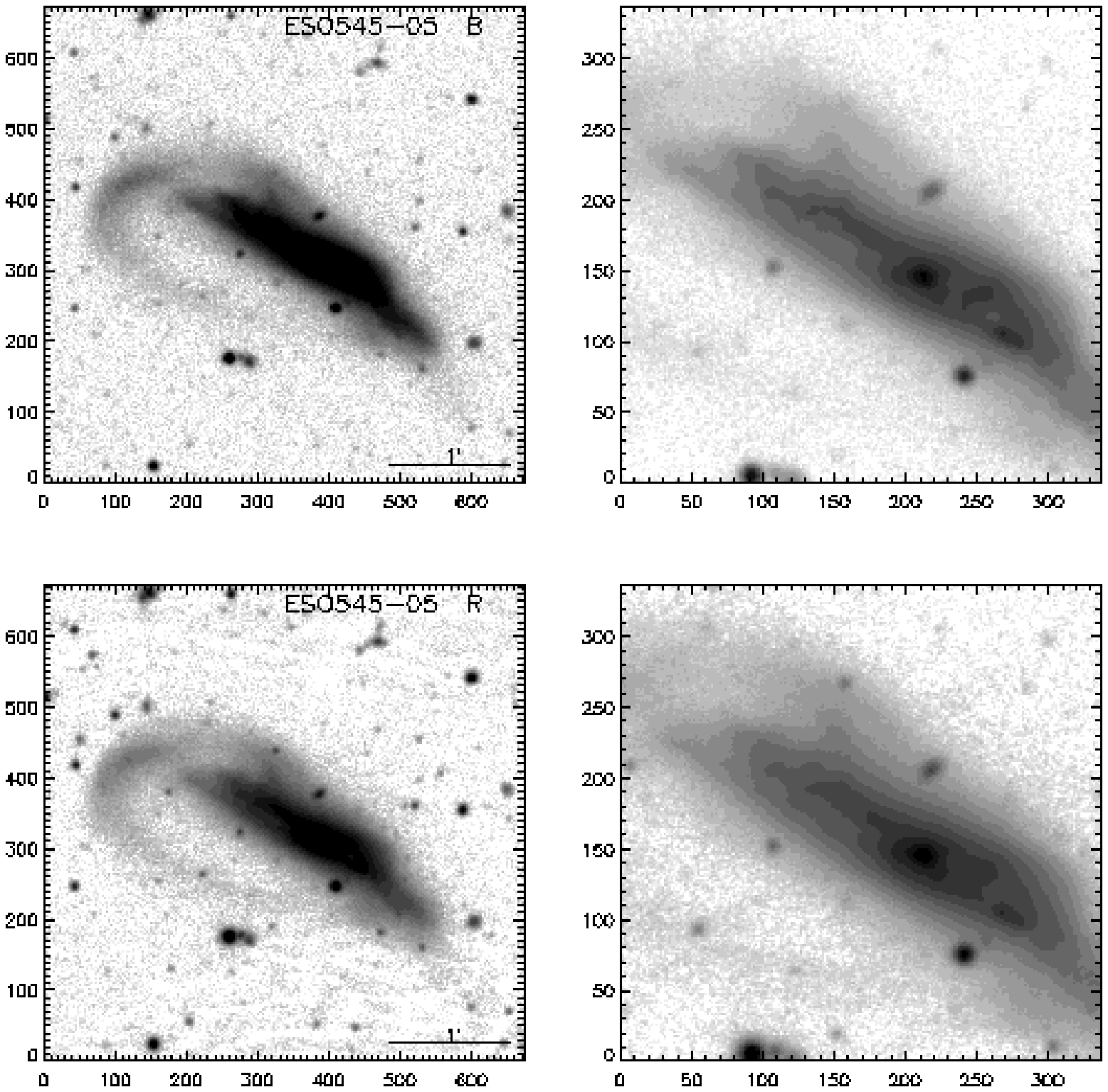}
\caption{Continued}
\end{figure}

\end{document}